\documentclass[superscriptaddress,dvipsnames,nofootinbib,amsmath,amssymb,aps,prd,floatfix,reprint]{revtex4-2}

\usepackage[titletoc,toc,title]{appendix}
\usepackage{mathtools}
\usepackage{amsmath, amsfonts,bm}
\usepackage[utf8]{inputenc}
\usepackage{xcolor}
\usepackage{graphicx}
\usepackage{float}
\usepackage{subfig}
\usepackage{comment}
\newcommand\myshade{85}
\colorlet{mylinkcolor}{RoyalPurple}
\colorlet{mycitecolor}{WildStrawberry}
\colorlet{myurlcolor}{BlueViolet}
\usepackage{orcidlink}

\usepackage[normalem]{ulem}

\usepackage{hyperref}
\usepackage[capitalise]{cleveref}

\hypersetup{
  linkcolor  = mylinkcolor!\myshade!black,
  citecolor  = mycitecolor!\myshade!black,
  urlcolor   = myurlcolor!\myshade!black,
  colorlinks = true,
}

\DeclareMathAlphabet{\mathup}{OT1}{\familydefault}{m}{n}

\newcommand{\be}{\begin{equation}} 
\newcommand{\ee}{\end{equation}}

\usepackage{array}
\newcommand{\PreserveBackslash}[1]{\let\temp=\\#1\let\\=\temp}
\newcolumntype{C}[1]{>{\PreserveBackslash\centering}p{#1}}
\newcolumntype{R}[1]{>{\PreserveBackslash\raggedleft}p{#1}}
\newcolumntype{L}[1]{>{\PreserveBackslash\raggedright}p{#1}}

\usepackage{multirow}
\usepackage{bm}

\begin{document}

\title{Impact of ACT DR6 and DESI DR2 for Early Dark Energy and the Hubble tension}

\author{Vivian Poulin\orcidlink{0000-0002-9117-5257}}
\email{vivian.poulin@umontpellier.fr}
\affiliation{Laboratoire univers et particules de Montpellier (LUPM), Centre national de la recherche scientifique (CNRS) et Universit\'e de Montpellier, Place Eug\`ene Bataillon, 34095 Montpellier C\'edex 05, France}
\author{Tristan L.~Smith\orcidlink{0000-0003-2685-5405}}
\email{tsmith2@swarthmore.edu}
\affiliation{Department of Physics and Astronomy, Swarthmore College, 500 College Ave., Swarthmore, PA 19081, USA}
\author{Rodrigo Calder\'on\orcidlink{0000-0002-8215-7292}}
\email{calderon@fzu.cz}
\affiliation{CEICO, Institute of Physics of the Czech Academy of Sciences, Na Slovance 1999/2, 182 21, Prague, Czech Republic}
\author{Th\'eo Simon\orcidlink{0000-0001-7858-6441}}
 \email{theo.simon@umontpellier.fr}
\affiliation{Laboratoire univers et particules de Montpellier (LUPM), Centre national de la recherche scientifique (CNRS) et Universit\'e de Montpellier, Place Eug\`ene Bataillon, 34095 Montpellier C\'edex 05, France}

\keywords{}

\begin{abstract}
The data release six of the Atacama Cosmology Telescope (ACT DR6) and the second data release from the Dark Energy Spectroscopic Instrument (DESI DR2) recently became available. In light of these data, we update constraints on the Early Dark Energy (EDE) resolution to the Hubble tension. While ACT DR6 does not favor EDE over the core cosmological model $\Lambda$CDM, it allows for a significantly larger maximum contribution of EDE, $f_{\rm EDE}$, in the pre-recombination era than the latest analysis of {\it Planck} NPIPE despite increased precision at small angular scales. Moreover, EDE rises the value of $H_0r_s$, improving consistency between CMB and DESI DR2 data. We find a residual tension with SH0ES of $\sim 2 \sigma$ for the combination of {\it Planck} at $\ell <1000$ + ACT DR6 + lensing + Pantheon-plus + DESI DR2, a significant decrease from $3.7 \sigma$ for analyses that use NPIPE and SDSS BAO data. A profile likelihood analysis reveals significant prior-volume effects in Bayesian analyses which do not include SH$0$ES, with confidence intervals of $f_{\rm EDE}=0.09\pm 0.03$ and $H_0= 71.0\pm1.1$ km/s/Mpc. When including DESI data, the EDE model with  $H_0=73$ km/s/Mpc provides a better fit than the $\Lambda$CDM model with $H_0=68.4$ km/s/Mpc. The inclusion of SH$0$ES data rises the preference well above $5\sigma$, with $\Delta\chi^2=-35.4$. Our work demonstrates that after ACT DR6 and DESI DR2, EDE remains a potential resolution to the Hubble tension.
\end{abstract}

\maketitle

\section{Introduction} \label{sec:intro}

Early Dark Energy (EDE) \cite{Karwal:2016vyq,Poulin:2018cxd,Smith:2019ihp} has emerged as a promising framework to resolve the Hubble tension between the value of the Hubble rate $H_0$ inferred from the core cosmological model, $\Lambda$-cold-dark-matter ($\Lambda$CDM),fit to Cosmic Microwave Background (CMB) data and that measured by calibrating type 1a supernovae (SN1a) by the SH0ES collaboration \cite{Riess:2021jrx} (see Refs.~\cite{Kamionkowski:2022pkx,Poulin:2024ken} for reviews). While the analyses of data from the Atacama Cosmology Telescope (ACT) data release 4 (DR4) had a slight preference for EDE over $\Lambda$CDM with no residual tensions with SH0ES~\cite{Hill:2021yec,Poulin:2021bjr,Smith:2022hwi}, analysis of EDE in light of the latest {\it Planck} NPIPE data-- Public Release 4 (PR4)-- \cite{NPIPE,Rosenberg:2022sdy,Tristram:2023haj} and baryonic acoustic oscillation (BAO) data from BOSS significantly reduced the success of this solution, only achieving a reduction to the $\sim 3.7\sigma$ level \cite{Efstathiou:2023fbn}. 

\begin{figure}
    \centering
    \includegraphics[width=1\linewidth]{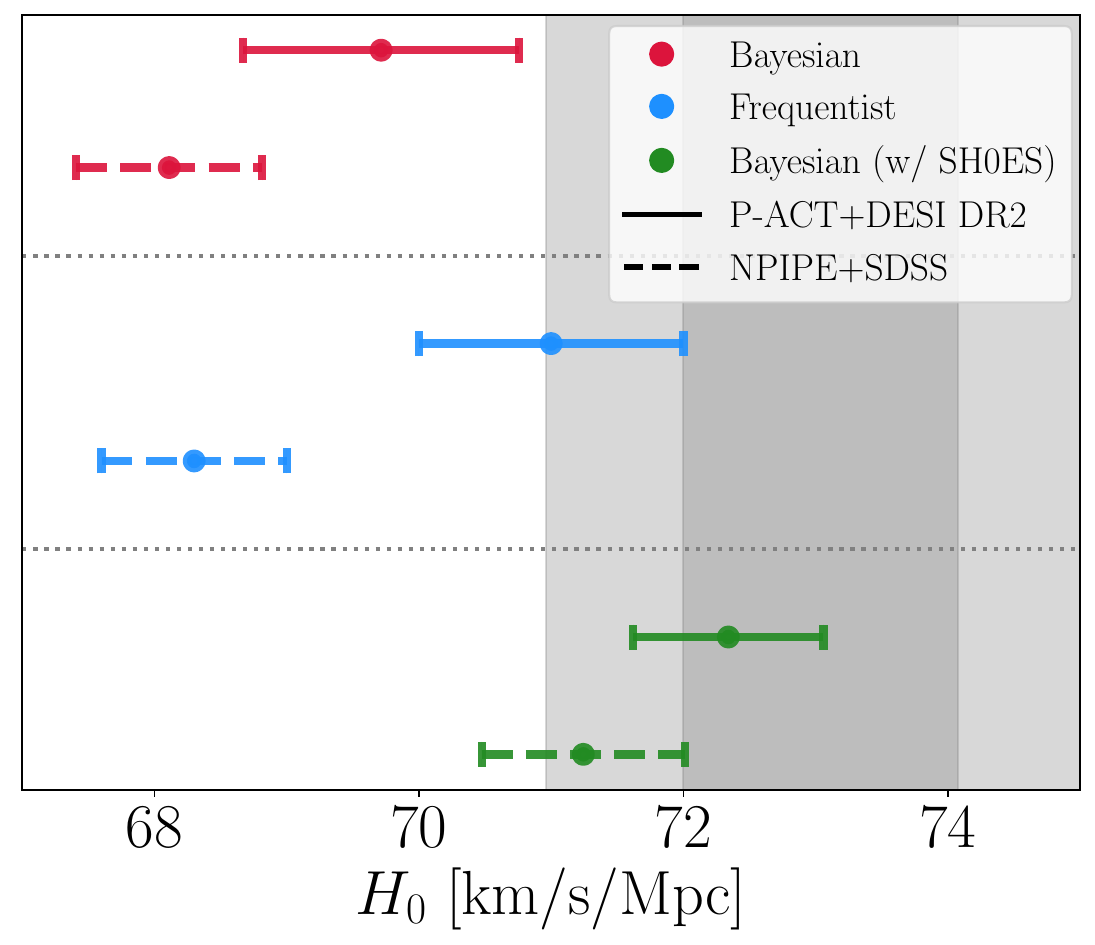}
    \caption{A whisker plot of the Hubble constant reconstructed from NPIPE+SDSS (from Ref.~\cite{Efstathiou:2023fbn})  and P-ACT+DESI DR2 (this work) in the `axion-like' EDE model. All analyses include CMB lensing and Pantheon-plus SN1a data. We quote credible and confidence intervals at 68\%.}
    \label{fig:summary}
\end{figure}

Recently, the ACT collaboration has determined new constraints on the axion-like EDE models from the combination of new DR6 data \cite{ACT:2025fju} with Planck public release 3 (PR3) \cite{Aghanim:2018eyx} restricted to multipoles $l < 1000$ in TT and  $l < 600$ in TE/EE   (dubbed `P-ACT' dataset), including the ACT DR6 CMB gravitational lensing \cite{ACT:2023kun} and updated BAO data from DESI DR1 \cite{DESI:2024mwx}. Contrary to previous analyses using ACT DR4 \cite{Hill:2021yec,Poulin:2021bjr,Smith:2002dz}, these data do not favor EDE over $\Lambda$CDM, resulting in upper limits to the maximum fractional EDE contribution to the total energy density, $f_{\rm EDE}(z_c)<0.12$ at the 95\% confidence level (CL) \cite{ACT:2025tim}. Yet, those constraints are significantly weaker than those coming from Planck NPIPE combined with SDSS BAO data, $f_{\rm EDE}(z_c)<0.061$ \cite{Efstathiou:2023fbn}. This is unexpected given that the same combination of data provide constraints that are as strong as NPIPE within $\Lambda$CDM \cite{ACT:2025fju}. 

In this paper, we quantify the ability of the EDE model to address the Hubble tension in light of the new ACT DR6 data and  Baryonic Acoustic Oscillation (BAO) data from the Dark Energy Spectroscopic Instrument (DESI) DR2 \cite{DESI:2025zgx}.\footnote{For a comparaison of ACT DR6 and DESI DR2 under $\Lambda$CDM, and the implications for dynamical dark energy and neutrino masses, we refer the reader to Ref.~\cite{DESI:2025gwf}.} We contrast those results with previous analyses using NPIPE and BAO data from the Sloan Digital Sky Survey (SDSS) \cite{Efstathiou:2024xcq}.  Our main results are presented in Fig.~\ref{fig:summary}, where we show the reconstructed Hubble constant from NPIPE and SDSS BAO data, and that from the P-ACT and DESI DR2 data, in the EDE model. All analyses further include the respective CMB lensing data and Pantheon-plus SN1a compilation \cite{Brout:2022vxf}.
In particular, we find that the newer data lead to less than a $2\sigma$ residual tension with SH$0$ES-- a strong decrease from what has been found in the latest pre-DESI and pre-ACT DR6 analysis, using NPIPE and SDSS BAO data in Ref.~\cite{Efstathiou:2023fbn}. Furthermore, we show that EDE provides better concordance between CMB and DESI BAO data than $\Lambda$CDM, in agreement with Ref.~\cite{chaussidon:2025}. Finally, we perform a profile likelihood analysis that reveals the extent to which prior volume effects affect the Bayesian analysis of the EDE model in light of ACT DR6 data when SH$0$ES data are left out of the analysis. The profile likelihood indicates that $f_{\rm EDE}(z_c) \neq 0$  at $\sim2.5\sigma$, and yields values of $H_0$ within $2\sigma$ of the SH$0$ES-inferred value.

The rest of the paper is structured as follows. After presenting our analysis setup in Sec.~\ref{sec:details}, we discuss in Sec.~\ref{sec:noBAO} the impact of the new ACT DR6 data at high-$\ell$. We then show the impact of the new DESI DR2 data compared to BOSS data in Sec.~\ref{sec:wBAO}. We discuss the profile likelihood analysis in Sec.~\ref{sec:profile} before concluding in Sec.~\ref{sec:concl}.

\section{Analysis setup}
\label{sec:details}

We focus here on the axion-like Early Dark Energy (EDE) model, characterized by a modified axion potential \cite{Kamionkowski:2014zda,Poulin:2018cxd,Smith:2019ihp}:  
\begin{equation}\label{eq:potential}
    V(\theta) = m^2 f^2[1-\cos (\theta)]^3,
\end{equation}  
where \( m \) denotes the axion mass, \( f \) represents the axion decay constant, and \( \theta \equiv \phi/f \) is a rescaled field variable constrained within \( -\pi \leq \theta \leq \pi \). The specific choice of `$[1-\cos \theta]^3$' leads to an energy density that redshifts faster than matter after the field becomes dynamical. This toy model is primarily chosen for its flexibility and simplicity, but it has been shown that it can be naturally embedded within string theory \cite{McDonough:2022pku,Cicoli:2023qri}. For alternative EDE parameterizations and reviews, we refer the reader to Refs.~\cite{Agrawal:2019lmo,Lin:2019qug,Alexander:2019rsc,Sakstein:2019fmf,Das:2020wfe,Niedermann:2019olb,Niedermann:2020dwg,Niedermann:2021vgd,Ye:2020btb,Berghaus:2019cls,Freese:2021rjq,Braglia:2020bym,Sabla:2021nfy,Sabla:2022xzj,Gomez-Valent:2021cbe,Moss:2021obd,Guendelman:2022cop,Karwal:2021vpk,McDonough:2021pdg,Wang:2022nap,Alexander:2022own,McDonough:2022pku,Nakagawa:2022knn,Gomez-Valent:2022bku,MohseniSadjadi:2022pfz,Kojima:2022fgo,Rudelius:2022gyu,Oikonomou:2020qah,Tian:2021omz,Maziashvili:2021mbm,Poulin:2023lkg,McDonough:2023qcu,Kamionkowski:2022pkx}.

We perform analyses using the public {\sc cobaya}~\cite{Torrado:2020dgo} package and analyze chains using {\sc getdist}~\cite{Lewis:2019xzd}. 
We consider combinations of the high-$\ell$ TTTEEE {\it Planck} PR4 CamSpec NPIPE likelihood \cite{NPIPE,2022arXiv220510869R} and low-$\ell$ TT (commander) and EE (sroll2) likelihoods \cite{Planck:2019nip,Pagano:2019tci},  the high-$\ell$ TTTEEE ACT DR6  \texttt{lite} data \cite{ACT:2025fju}, the {\it Planck} and ACT DR6 lensing likelihoods \cite{Planck:2018vyg,ACT:2023kun}, BAO from DESI DR2 \cite{DESI:2024mwx,DESI:2025zgx} and the Pantheon-plus catalog \cite{Brout:2022vxf}.
When combining {\it Planck} and ACT DR6 we follow the ACT collaboration's recommendation and impose a cut in the multipoles of the various {\it Planck} likelihoods.\footnote{We use the official `{\sc PlanckActCut}' likelihood available in {\sc cobaya}, based on PR3} We refer to {\it Planck} data as `NPIPE' for the full likelihood and simply as `P' for the cut likelihood, while we refer to ACT DR6 data as `ACT', lensing data as `L', BAO data as `B' and Pantheon-plus SN1a data as `S'.
In addition, the SH0ES Cepheid calibration of the peak SN1a absolute magnitude is modeled as a Gaussian,  $M_b=-19.253 \pm
0.027$~\cite{Riess:2021jrx}. We refer to it as `$M_b$'. To assess consistency of the various datasets with SH$0$ES, we make use of the `difference in maximum a posteriori' (DMAP) statistics \cite{Raveri:2019}, $Q_{\rm DMAP} \equiv \sqrt{\chi_{\rm tot}^2({\rm with ~SH0ES})-\chi_{\rm tot}^2({\rm without~SH0ES})}$, where the two $\chi^2$ values
are computed at the maximum a posteriori points with and without including the SH$0$ES $M_b$ prior. To assess preference\footnote{As EDE has three extra free parameters compared to $\Lambda$CDM, one can compute the $\Delta$AIC$=\Delta\chi^2+6$ to quickly gauge the preference over $\Lambda$CDM \cite{Akaike}. } over $\Lambda$CDM we report the $\Delta\chi^2\equiv \chi_{\rm tot}^2({\rm EDE})-\chi_{\rm tot}^2({\rm\Lambda CDM})$. 

To make the Bayesian inference numerically tractable, we make use of the emulator of the EDE model\footnote{The emulator further includes 3 neutrinos of degenerate mass wich $\sum m_\nu =0.06$ eV.} described in Ref.~\cite{Qu:2024lpx}, as is done by the ACT collaboration. We run Monte Carlo Markov chains (MCMC) using the metropolis-hasting algorithm and consider chains to be converged when they fulfill the Gelman-Rubin criterion $R-1 < 0.01$.  We take as free parameters the standard six $\Lambda$CDM parameters (varied within large flat priors), namely the (dimensionless) Hubble rate $h\equiv H_0/(100{\rm km/s/Mpc})$, the baryon density $\omega_b\equiv\Omega_bh^2$, the cold dark matter density $\omega_{\rm cdm}\equiv \Omega_{\rm cdm}h^2$, the optical depth to reionization $\tau_{\rm reio}$, as well as the primordial power spectrum tilt $n_s$ and amplitude $A_s$. We also vary  the critical redshift below which the field dilutes $z_c\in[3,4.3]$, the associated EDE fraction corresponding to the maximum contribution  $f_{\rm EDE}(z_c)\in[0.001,0.5]$ and the initial field value $\theta_i \equiv \phi_i/f\in [0.1,3.1]$. We provide Bayesian credible intervals of the cosmological parameters for all analyses performed in this work in App.~\ref{app:tables}, Tab.~\ref{tab:CI}.

When computing the best fit $\chi^2$, we use the {\sc prospect} package \cite{Holm:2023uwa}.\footnote{When computing the best fit $\chi^2$, we follow recommendation of the ACT collaboration and make use of the code \texttt{class\_ede} \cite{Hill:2020osr} instead of the emulator. In that case, we follow the ACT collaboration and use a single massive neutrinos of mass $m_\nu=0.06$ eV and two massless ones for both $\Lambda$CDM and EDE.} Tables listing individual data $\chi^2_{\rm min}$ are provided in App.~\ref{app:tables}.
In addition, for some data combinations, we also use {\sc prospect} to compute the profile likelihood\footnote{In that case, we use the emulator of the EDE model given the long computational time of \texttt{class\_ede} with high accuracy precision parameters. We compare the use of the full code on some specific points in App.~\ref{app:profile} finding reasonable agreements.} of the maximal EDE fraction $f_{\rm EDE}$ and the Hubble rate $H_0$.   This allows us to contrast the Bayesian results that are affected by prior volume effects \cite{Herold:2021ksg,Herold:2022iib,Karwal:2024qpt} to illustrate the true constraining power of the data likelihoods. 

\section{Impact of ACT DR6 for EDE and the Hubble tension}
\label{sec:noBAO}

\begin{figure*}[!t]
    \centering
    \includegraphics[width=0.49\linewidth]{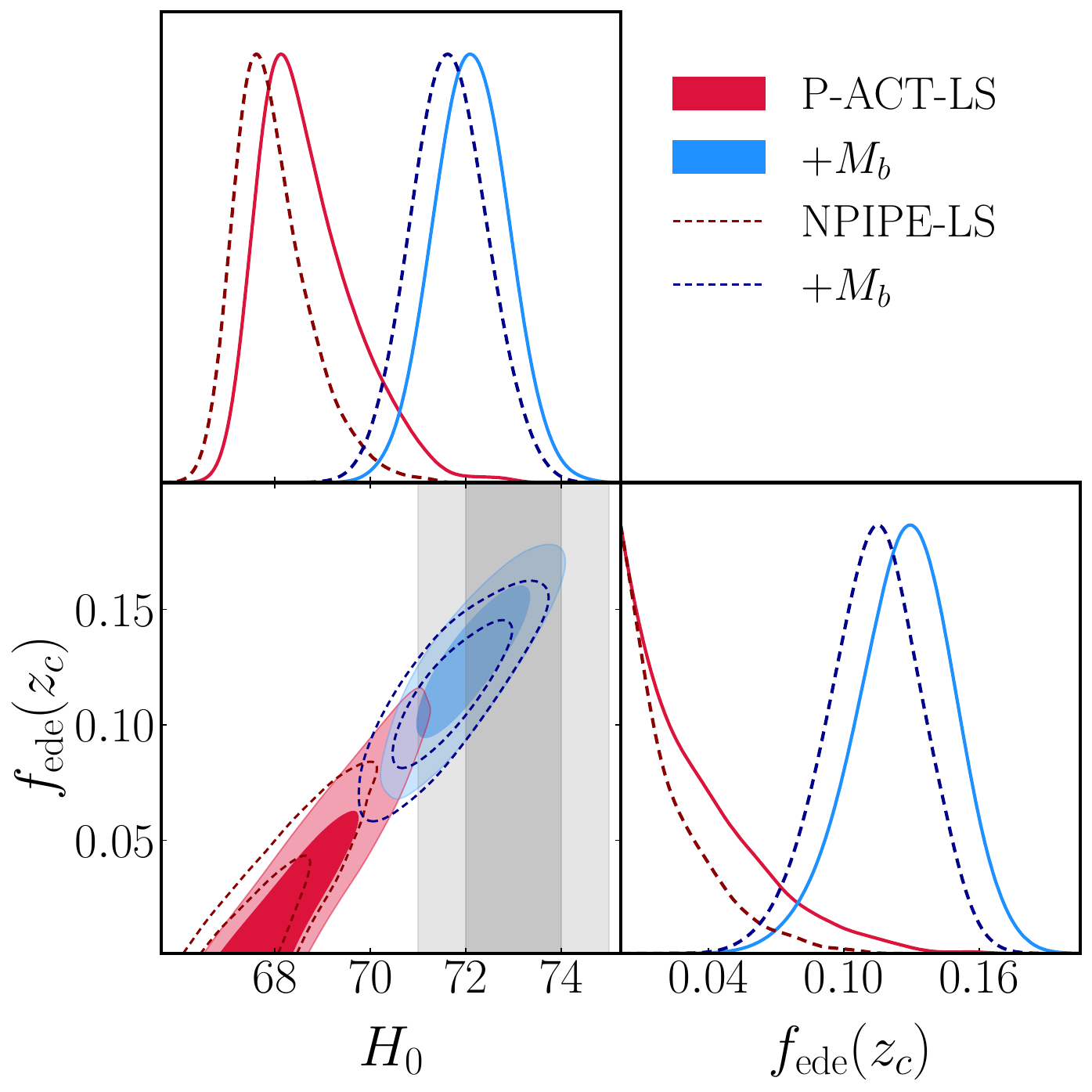}
    \includegraphics[width=0.49\linewidth]{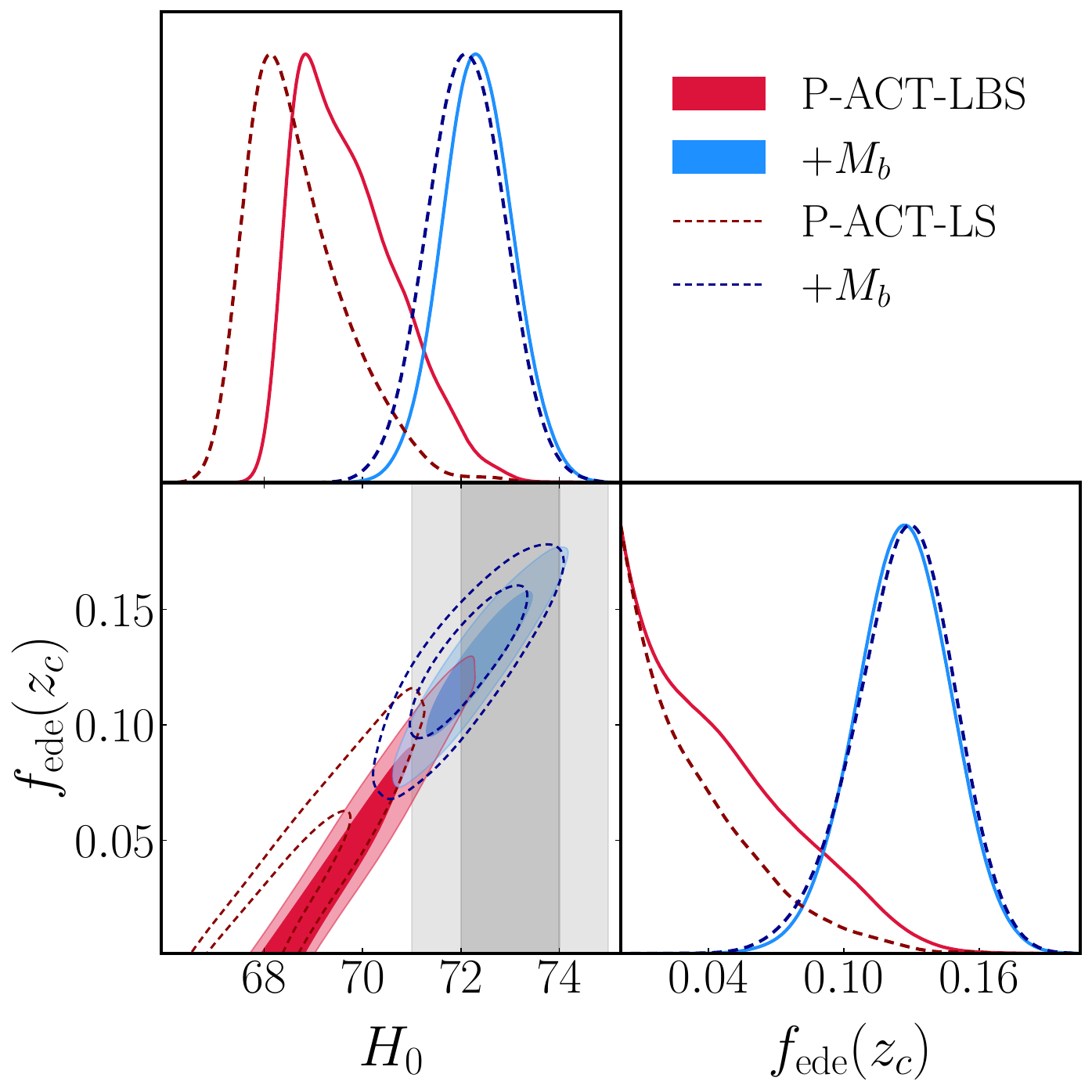}
    \caption{{\it left panel:} 1D and 2D posterior distributions of $f_{\rm EDE}$ and $H_0$ resulting from the analyses of either {\it Planck} (NPIPE) or  {\it Planck}+ACT DR6 (P-ACT), together with lensing (L) and Pantheon-plus (S) data, both with and without SH$0$ES ($M_b$). {\it right panel:} Same as left panel, now comparing the use of DESI BAO data (B) in the P-ACT analysis. }
    \label{fig:NPIPE-vs-ACT-vs-DESI}
\end{figure*}

We start by comparing the constraints from {\it Planck} NPIPE to those using the ACT DR6 data. In the left panel of Fig.~\ref{fig:NPIPE-vs-ACT-vs-DESI} we show the 1D and 2D posterior distributions of $f_{\rm EDE}$ and $H_0$ resulting from the analyses of either {\it Planck} NPIPE or  {\it Planck}+ACT DR6, together with lensing and Pantheon-plus data (both with and without SH$0$ES).
In this first analysis, we do not include BAO data to gauge the sole impact of the ACT DR6 CMB data. It is striking that the {\it Planck}+ACT DR6 yields significantly weaker constraints than the {\it Planck} NPIPE analysis. In the analyses without the SH$0$ES prior we find roughly 40\% weaker constraints and a $\gtrsim1\sigma$ shift to $h$:
\begin{eqnarray}
\nonumber
f_{\rm EDE}(z_c) & < &  0.065,~100h=67.96^{+0.51}_{-0.93}~~(\mathrm{NPIPE\text{-}LS})\\
\nonumber
f_{\rm EDE}(z_c) & <  &0.092,~100h=68.68^{+0.62}_{-1.2}~~(\mathrm{P\text{-}ACT\text{-}LS})\,,
\end{eqnarray}
where $h \equiv H_0/(100\ {\rm km/s/Mpc})$. Upper limits are provided at the 95\% confidence level (CL) while two-sided bounds are provided at 68\% C.L.
Note that under $\Lambda$CDM the constraints from  P-ACT are as strong as those from NPIPE (and $20\%$ stronger than {\it Plik} PR3 \cite{Aghanim:2018eyx}); hence it is not trivial that constraints to EDE get significantly weaker. 
The preference for EDE in those analyses is negligible for NPIPE ($\Delta \chi^2=-0.9)$ and small for P-ACT ($\Delta \chi^2=-5.0)$. 
Furthermore, we find that the inclusion of Pantheon-plus appears to strengthen the Bayesian constraints on $f_{\rm EDE}$ compared to what is found by the ACT collaboration without those data ($f_{\rm EDE} <0.12$) \cite{ACT:2025tim}.  

The tension with SH$0$ES computed with the  $Q_{\rm DMAP}$ metric (which captures the non-Gaussianity of the posteriors) is significantly weaker for  P-ACT-LS ($2.6\sigma$) than for NPIPE-LS ($3.5\sigma$). For comparison, under $\Lambda$CDM the same metric yields a $6.2\sigma$ tension for  P-ACT-LS and $6.5\sigma$ tension for NPIPE-LS. The residual level of tension is similar to that obtained with {\it Plik} PR3 \cite{Smith:2023oop}, despite larger statistical power in the P-ACT analysis.

When combining with SH$0$ES, we find 
\begin{eqnarray}
\nonumber
f_{\rm EDE}(z_c) =  0.113^{+0.022}_{-0.020},~100h=71.65\pm0.81&~~\\
\nonumber(\mathrm{NPIPE\text{-}LSM}_b)&\,,\\
\nonumber
f_{\rm EDE}(z_c) =  0.127^{+0.024}_{-0.020},~100h=72.11\pm0.79&\\\nonumber
(\mathrm{P\text{-}ACT\text{-}LSM}_b)&\,.
\end{eqnarray}
The inclusion of ACT DR6 thus increases the mean value of $f_{\rm EDE}(z_c)$ and $h$ by $\sim 0.5\sigma$, with detection of EDE that exceeds the $6.3\sigma$  level, leading to a strong preference over $\Lambda$CDM ($\Delta\chi^2=-30.7$). 

\section{A comparison with ACT DR4 and NPIPE}

\begin{figure*}[!t]
    \centering
    \includegraphics[width=2\columnwidth]{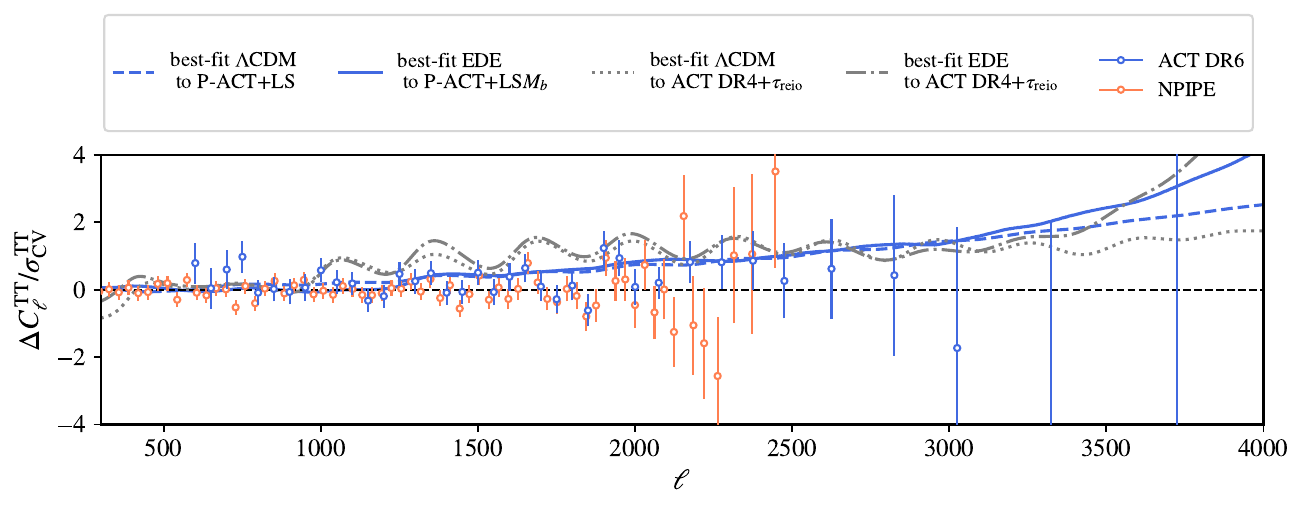}
    \caption{The best fit $\Lambda$CDM and EDE residuals with respect to the best fit $\Lambda$CDM model to NPIPE  \cite{Rosenberg:2022sdy} normalized by the cosmic variance (CV) limits for the temperature power spectrum (note that the data may have uncertainties smaller than the CV limit since they are binned). We show residuals for the best fit $\Lambda$CDM model using P-ACT+LS (dashed), the best fit $\Lambda$CDM model using ACT DR4 + a prior on $\tau_{\rm reio}$ (dotted), while for EDE we use P-ACT+LS$M_b$ (solid) and ACT DR4 + a prior on $\tau_{\rm reio}$ (dot dashed). The ACT DR4 residuals were computed using the best fit parameters given in Ref.~\cite{Hill:2021yec}. The ACT DR6 bandpowers are marginalized over foreground and other nuisance parameters (i.e. the `lite' versions of their respective likelihoods) \cite{ACT:2020gnv,ACT:2025fju}, while the NPIPE bandpowers are computed at the best fit foreground model (from Ref.~\cite{Efstathiou:2023fbn}). We note that these bandpowers are not statistically independent, especially at $\ell \gtrsim 2000$ where the foreground model leads to strong correlations. }
    \label{fig:residualsNPIPE}
\end{figure*}
The analysis of the previous ACT data release-- DR4-- showed tantalizing evidence for EDE ($f_{\rm EDE}(z_c) = 0.142^{+0.039}_{-0.072}$) along with a significant increase in the mean value of $H_0$ ($H_0 = 74.5^{+2.5}_{-4.4}$ km/s/Mpc)
without including any other data sets \cite{Hill:2021yec,Poulin:2021bjr}. Since ACT DR6 now places an upper limit on $f_{\rm EDE}$ despite increase statistical power, one might conclude that the new data `rule out' EDE as a resolution to the Hubble tension.

As we now argue, the shift in the EDE constraints from DR4 to DR6 is not surprising given that the \emph{Planck} and South Pole Telescope (SPT) measurements \cite{SPT-3G:2022hvq} of the CMB are in significant tension with the best-fit EDE cosmology to ACT DR4 \cite{Efstathiou:2023fbn,Smith:2023oop}. To illustrate this, we show a comparison of the residuals of the best-fit $\Lambda$CDM and EDE models to either P-ACT-LS (+$M_B$ for EDE) or ACT DR4 and a prior on $\tau_{\rm reio}$ (with parameters taken from Ref.~\cite{Hill:2021yec}), as well as the NPIPE and DR6 bandpowers in Fig.~\ref{fig:residualsNPIPE}. 
These residuals have been computed with respect to the NPIPE $\Lambda$CDM best fit model \cite{Rosenberg:2022sdy}.  
A similar figure, explicitly comparing ACT DR6 with ACT DR4 data residuals, is provided in App.~\ref{app:DR4}.

For all models, the residuals show an enhancement of power with respect to the NPIPE $\Lambda$CDM best fit model as we go to smaller angular scales. This trend is directly related to ACT's preference for a larger scalar spectral index\footnote{This is also consistent with constraints on the number of effective neutrino species found for ACT DR4, namely $N_{\rm eff} = 2.42 \pm0.41$ \cite{ACT:2020gnv}, and P-ACT DR6, namely $N_{\rm eff} = 2.73 \pm0.14$ \cite{ACT:2025tim}. This $\sim 2 \sigma$ preference for $N_{\rm eff} < 3$ leads to an increase in the photon diffusion damping scale with, in turn, an enhancement in power at small scales. }, $n_s$, compared to \emph{Planck}-- for DR4 the best fit $\Lambda$CDM value is $n_s = 1.013$ and for P-ACT DR6 it is $n_s = 0.9709$, to be compared with NPIPE's best fit value of $n_s = 0.9635$. 

As was previously shown in Ref.~\cite{Efstathiou:2023fbn} it is clear that the best fit to ACT DR4 +$\tau_{\rm reio}$ is strongly excluded by NPIPE data, in particular due to much larger power at $\ell \sim 1000-2000$.
If the ACT DR4 best fit EDE model continued to be favored by DR6 then there would be a strong tension with \emph{Planck} data. BOSS data analyzed under the effective field theory of large scale structure would also strongly exclude this model \cite{Hill:2020osr,DAmico:2020ods,Simon:2022adh}.

The fact that the ACT DR6 data has a lower amplitude than ACT DR4 at large multipoles reduces the tension with NPIPE to $\sim 2.6 \sigma$, as quantified by the ACT collaboration \cite{ACT:2025fju}.
These differences between ACT DR6 and NPIPE could be due to statistical fluctuations, but they have important implications for models involved in resolving the Hubble tension. 
The next data release of the SPT collaboration will be important to clarify the situation.

\section{Impact of DESI DR2: the role of $H_0r_s$}
\label{sec:wBAO}
We now turn to including DESI DR2 data in the analysis. For analyses that use DESI DR1 data, we refer the reader to Refs.~\cite{Qu:2024lpx,Poulin:2024ken}. 
We show in the right panel of Fig.~\ref{fig:NPIPE-vs-ACT-vs-DESI},  the 1D and 2D posterior distributions from the analyses of P-ACT with and without DESI DR2 data (always including other datasets as described above). The inclusion of DESI DR2 data {\it relaxes} the constraints to $f_{\rm EDE}(z_c)$ by $\sim 20 \%$ and further rises the value of $h$, and we find:
\begin{eqnarray}
\nonumber
f_{\rm EDE}(z_c) <  0.109,~100h=69.71^{+0.64}_{-1.3}&~~\\
\nonumber(\mathrm{P\text{-}ACT\text{-}LBS})&\,,\\
\nonumber
f_{\rm EDE}(z_c)=  0.126\pm0.021,~100h=72.34\pm0.72&\\\nonumber
(\mathrm{P\text{-}ACT\text{-}LBSM}_b)&\,.
\end{eqnarray}
The relaxation of the constraints is also reflected in the $\chi^2$-statistics, as we find $\Delta \chi^2=-7.0$ (without SH$0$ES) and $\Delta \chi^2=-35.4$ (with SH$0$ES) and a $Q_{\rm DMAP}$-tension metric of $2.0\sigma$. 
Comparing our results to those from the ACT collaboration we find that they are in good agreement: we find similar a $\Delta \chi^2$, though our constraints on $f_{\rm EDE}(z_c)$ are $\sim 10\%$ stronger. This slight difference is likely due to the inclusion of Pantheon-plus and DESI DR2 (instead of DESI DR1) in our analysis. 

It is also instructive to compare our results to the latest analysis pre-DESI and pre-ACT DR6 data, performed in Ref.~\cite{Efstathiou:2023fbn}, where $H_0$ constraints are displayed in Fig.~\ref{fig:summary}. This analysis includes data from {\it Planck} NPIPE (and the low-$\ell$ TT, EE and lensing likelihoods from Planck PR3 \cite{Aghanim:2018eyx}), BAO from SDSS \cite{eBOSS:2020lta,Reid:2015gra} and SN1a from Pantheon-plus. Therein it was found $f_{\rm EDE}(z_c)<0.061$ and $100h=68.11^{+0.47}_{-0.82}$, in $3.7\sigma$ tension with SH$0$ES. Hence, the update of CMB data (including lensing) from NPIPE to P-ACT and BAO from SDSS to DESI DR2 has improved the ability of the EDE model to reduce the tension by almost $2\sigma$.
The stark contrast between the strength of the constraints coming from NPIPE+SDSS and those from P-ACT-LBS, and the implication for the Hubble tension, warrants further study to clarify the origin of the difference. Similar results were found when comparing ACT DR4 data with \texttt{plik} in Ref.~\cite{Smith:2022hwi}.

\begin{figure}
    \centering
    \includegraphics[width=1\columnwidth]{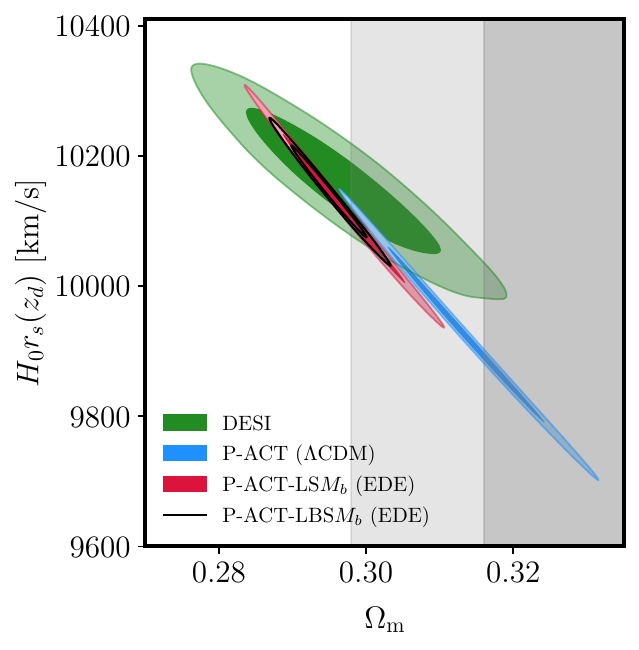}
    \caption{ 2D posterior distribution of $\Omega_m$-vs-$H_0r_s$ reconstructed in various analyses of EDE and $\Lambda$CDM. We compare in particular the reconstruction from P-ACT under $\Lambda$CDM and from P-ACT-LS$M_b$ under EDE to the value favored by DESI alone (that is identical under both EDE and $\Lambda$CDM). We also show the Pantheon-plus SN1a  $\Omega_m$ constraint in gray.}
    \label{fig:H0rs-vs-Omegam}
\end{figure}

\begin{figure*}[!t]
    \centering
    \includegraphics[width=1\columnwidth]{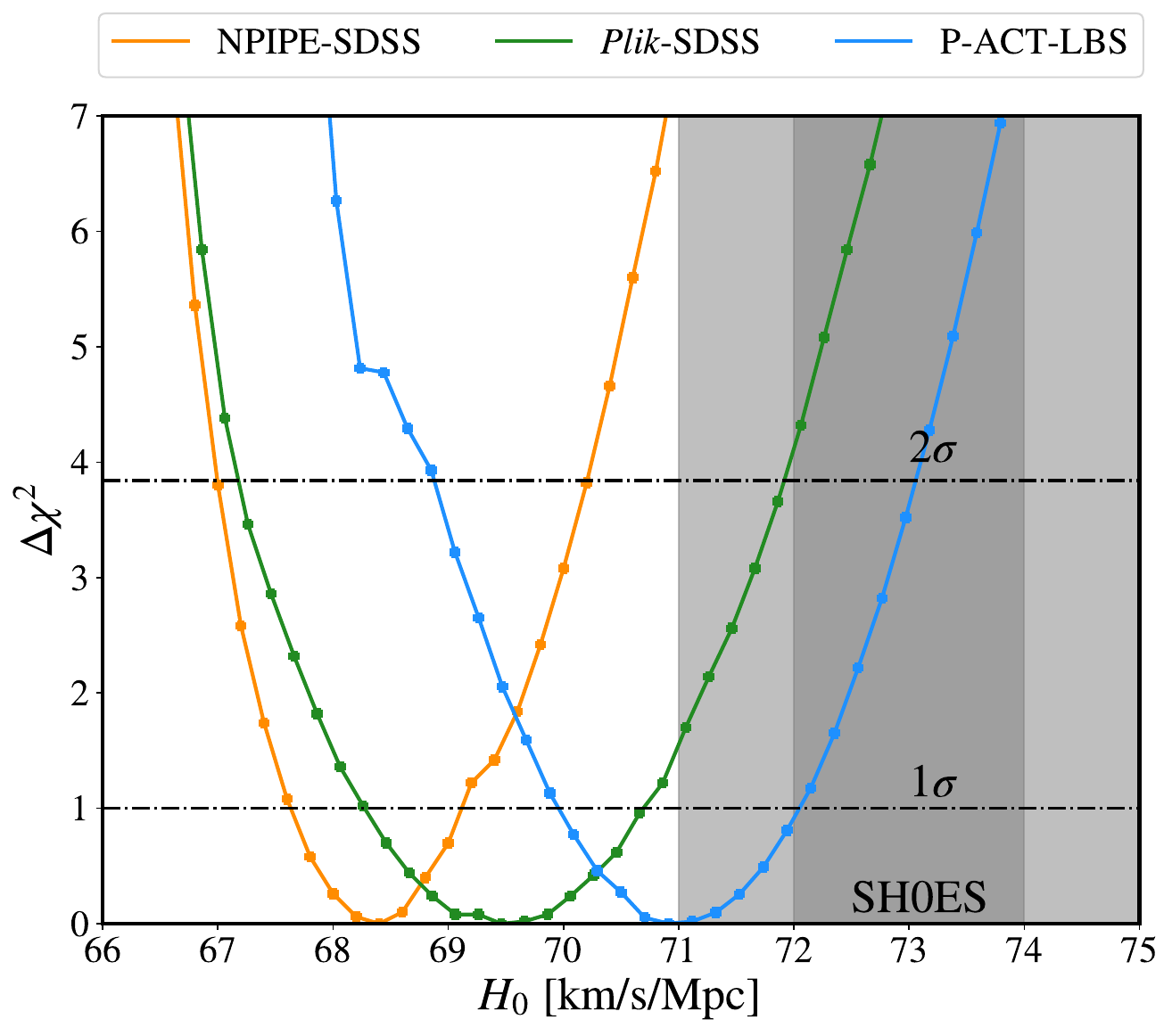}
        \includegraphics[width=1\columnwidth]{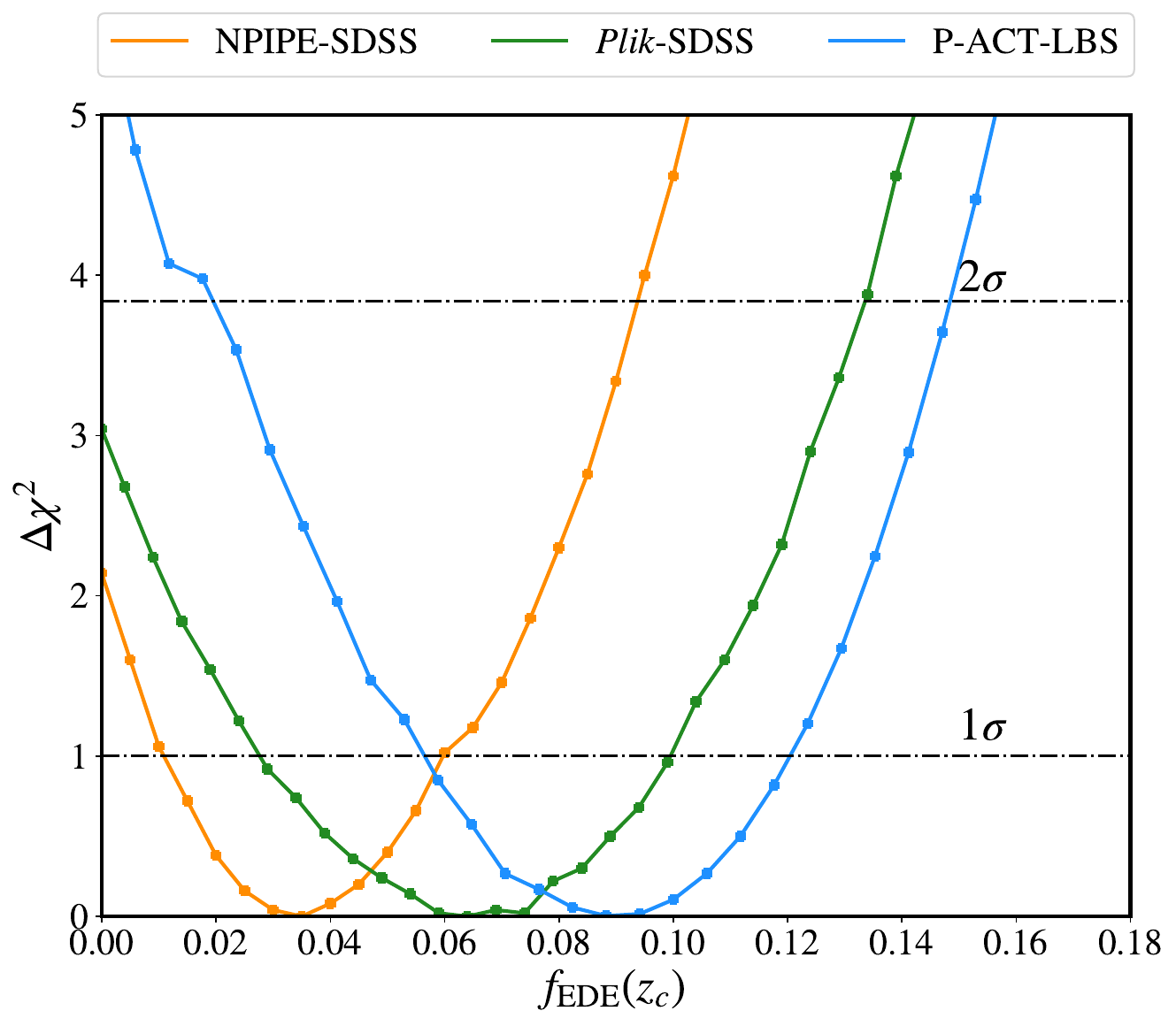}

    \caption{Profile likelihood of $H_0$ (left panel) and $f_{\rm EDE}(z_c)$ (right panel) for NPIPE+SDSS, {\it Plik}+SDSS and P-ACT-DESI. All analyses include Pantheon-plus (uncalibrated) and lensing data.}
    \label{fig:profile}
\end{figure*}

To understand the role of DESI DR2, we show in Fig.~\ref{fig:H0rs-vs-Omegam} the 2D posterior distribution of $\Omega_m$-vs-$H_0r_s$ reconstructed in various analyses of EDE and $\Lambda$CDM. We compare in particular the reconstruction from P-ACT under $\Lambda$CDM and from P-ACT-LS$M_b$ under EDE to the value favored by DESI alone (that is identical under both EDE and $\Lambda$CDM). We also show the Pantheon-plus SN1a  $\Omega_m$ constraint. 
One can see the known `$H_0r_s$-$\Omega_m$ tension' between CMB and DESI BAO, and the `$\Omega_m$ tension' between BAO and SN1a, that leads to a preference for dynamical dark energy \cite{Tang:2024lmo, DESI:2025zgx}. 
Interestingly, the EDE cosmology that fits SH$0$ES predicts a larger value for $H_0r_s$ and a slightly smaller $
\Omega_m$ in good agreement with DESI (see also Refs.~\cite{chaussidon:2025,Pang:2025lvh,Lynch:2024hzh,Mirpoorian:2025rfp}). 
As a result, the inclusion of DESI DR2 data in the analyses increases the preference for EDE.
Indeed, while the role of EDE is to decrease $r_s$, the dark matter density $\omega_{\rm c}\equiv\Omega_ch^2$ increases in the EDE cosmology to preserve the fit to the CMB \cite{Vagnozzi:2021gjh,Poulin:2024ken}.
This impacts the angular diameter distance, necessitating a larger increase in $H_0$ than a decrease in $r_s$. 

However, the `$\Omega_m$ tension' is slightly exacerbated as $\Omega_m$ in the EDE cosmology\footnote{The analyses we report here include Pantheon-plus, and thus one should be careful in comparing this value of $\Omega_m$ with that inferred from analyzing Pantheon-plus under $\Lambda$CDM as it double counts information. We checked that leaving Pantheon-plus out of the analysis, with SH$0$ES included as a prior on $H_0$, as a negligible impact on the value of $\Omega_m$.} is measured to be $\Omega_m= 0.2949\pm0.0034$. This is in $2.1\sigma$ tension with the value inferred from Pantheon-plus and $3.1\sigma$ tension with that from DESY5 \cite{DES:2024tys}. This result suggests that EDE cannot fully remove the preference for dynamical dark energy in analyses that combine CMB, BAO and SN1a. This was explicitly shown with DESI DR1 data in previous work \cite{Poulin:2024ken} and is supported by results that use NPIPE and DESI DR2 \cite{chaussidon:2025,Pang:2025lvh}. Though we note that Ref.~\cite{Poulin:2024ken} also found that the tension with SH$0$ES remains unchanged when letting the equation of state of DE free to vary. 
We leave a dedicated study of EDE and dynamical dark energy with ACT DR6 to future work, as the emulator we use \cite{Qu:2024lpx} was not trained on a cosmology that leaves the DE equation of state free to vary.

\section{Profile likelihood analysis of ACT DR6 and DESI data}
\label{sec:profile}

We now investigate the impact of prior volume effects in the analysis that does not include a SH$0$ES prior (i.e., that the posterior distributions are significantly affected by our choice of priors \cite{Smith:2020rxx}). We follow Refs.~\cite{Herold:2021ksg,Herold:2022iib} and perform a profile likelihood analysis using {\sc prospect}, shown in Fig.~\ref{fig:profile}. We reproduce the results presented in Ref.~\cite{Efstathiou:2023fbn}, that perform the same exercise with NPIPE+SDSS and {\it Plik}+SDSS (also including Pantheon-plus and {\it Plik} lensing data). 
 We build confidence intervals from the profile using the Neyman construction~\cite{Neyman:1937uhy}, which for 1D profile amounts in computing $\Delta\chi^2<1~(3.84)$ at 68\% (95\%) C.L. This procedure yields:
\begin{equation}
    f_{\rm EDE}(z_c)=  0.090^{+0.030}_{-0.033},~~100h = 71.96^{+1.08}_{-0.99}\,.\nonumber
\end{equation}

These intervals are significantly different than their Bayesian counterparts. In particular, one can see that in a frequentist analysis, $f_{\rm EDE}(z_c)>0$ at $\sim 2.5\sigma$ in the P-ACT-LBS analysis\footnote{Note that the profile computed with the emulator shows a small non-gaussianity towards low $f_{\rm EDE}(z_c)$. See App.~\ref{app:profile} for a comparison with the full code.}, while the Bayesian analysis yielded only an upper limit on the EDE fraction. We also find a significantly higher value of the Hubble rate in the frequentist analysis obtained without information from SH$0$ES. This demonstrates that analyses that include P-ACT are affected by prior volume effects: the apparent constraint to $f_{\rm EDE}$  derived in a Bayesian analysis is not driven by a degradation of the fit to the data, but by the large fraction of prior volume where the model is indistinguishable from $\Lambda$CDM and is thus disfavored due to unnecessary additional parameters. We stress that this is a perfectly valid concern (the region of parameter space that can address the tension is small), but it should be understood as an important caveat to the Bayesian constraints. 
The profile further illustrates the impact of updating CMB and BAO data, as the confidence intervals shift by more than $3\sigma$ compared to the NPIPE-SDSS analysis and about 1$\sigma$ compared to the {\it Plik}-SDSS analysis. 
This also corroborates the results of the ACT collaboration, which found significantly better agreement with {\it Plik} data than NPIPE data.

Comparing the EDE and $\Lambda$CDM fits, the $\chi^2$ values presented in Tables~\ref{tab:LCDM_chi2} and \ref{tab:EDE_chi2} (see Appendix~\ref{app:chi2}) demonstrate that EDE consistently provides a better fit to ACT data than $\Lambda$CDM across all analyses performed in this work. However, when comparing different EDE fits, we observe a slight degradation in the ACT $\chi^2$ within the combined analyses that include SH0ES and DESI data, relative to the EDE analysis excluding those datasets (i.e., comparing P-ACT-LS to P-ACT-LS$M_b$ and P-ACT-LBS$M_b$ within the EDE framework). Specifically, the ACT $\chi^2$ increases by 2 in P-ACT-LS$M_b$ and by 3 in P-ACT-LBS$M_b$. On the other hand, the fit to DESI data is improved (even under $\Lambda$CDM) in the analyses that include SH$0$ES. Overall, we find that when including DESI data, the EDE model with  $H_0=73$ km/s/Mpc provides a better fit by $\Delta\chi^2= -3.0$ to P-ACT-LBS than the $\Lambda$CDM model with $H_0=68.4$ km/s/Mpc.
Without DESI, the value $H_0=73$ km/s/Mpc in EDE is a worse fit by $\Delta\chi^2= +2.5$ than the $\Lambda$CDM fit.
For comparison, with NPIPE and SDSS BAO, the overall fit is worsened by $\Delta\chi^2=+14.3$ \cite{Efstathiou:2023fbn}. This confirms that, while ACT data do not favor EDE, they are much more permissive to large EDE fraction and large $H_0$ values. 
 
\section{Summary and Conclusions}

In this paper, we have shown that the combination of ACT DR6 data with Planck PR3 (with the TT spectrum restricted to $\ell \leq 1000$ and the EE/TE spectra restricted to $\ell \leq 600$) and updated BAO measurements from DESI DR2 yields constraints on axion-like EDE models that are $\sim 40\%$ weaker than those from NPIPE combined with SDSS BAO data, despite providing comparable constraints within $\Lambda$CDM. The updated data combination leads to less than $2\sigma$ tension with the SH$0$ES determination of the Hubble constant, significantly reduced compared to the $\sim 3.7\sigma$ tension seen with NPIPE and SDSS BAO data \cite{Efstathiou:2024xcq}. The origin of the differences, which could be statistical fluctuations in either or both datasets, deserves further investigation. We further demonstrate that EDE improves the agreement between the CMB and DESI DR2 BAO datasets relative to $\Lambda$CDM as it rises the value of $H_0r_s$, supporting earlier findings with NPIPE \cite{chaussidon:2025}. However, the value of  $\Omega_m$ in the EDE cosmology that fits SH0ES is in 2$\sigma$ and $3\sigma$ tension with that inferred from the Pantheon-plus SN1a and DES-Y5 catalog, respectively. We anticipate that the tension would be resolved if one also considers letting the equation of state of dark energy free to vary, as found in Ref.~\cite{Poulin:2024ken,Pang:2025lvh} with NPIPE data, though this remains to be demonstrated and the impact for the Hubble tension investigated. 

Finally, we have performed a profile likelihood analysis, demonstrating that the Bayesian upper limits in the P-ACT analysis on $f_{\rm EDE}$ are strongly influenced by the volume of prior space, with the frequentist analysis (i.e., goodness of fit) yielding a nonzero EDE contribution at $\sim2.5\sigma$.  In fact, when DESI is included in the analysis, the EDE model with $H_0=73$ km/s/Mpc provides a better fit by $\Delta \chi^2 =-3.0$ to the data than the $\Lambda$CDM model with $H_0=68.4$ km/s/Mpc.
These results support the conclusion that EDE remains a potential resolution to the Hubble tension after the release of ACT DR6 and DESI DR2 data.

An enhancement in power at small scales is a generic prediction of EDE models (see, e.g., Refs.~\cite{Smith:2022hwi,Poulin:2023lkg,Smith:2025arq}). It is also important to note that not all models that attempt to resolve the Hubble tension predict this enhancement. For example, Wess-Zumino dark radiation (WZDR) model \cite{Aloni:2021eaq} predicts a decrement so that we expect the ACT DR6 data will place stronger limits on this model \cite{Smith:2025arq}. Upper limits to the effective number of strongly interacting dark radiation (i.e. perfect fluid dark radiation)-- which captures some of the key features of WZDR-- decreases to $N_{\rm idr} < 0.114$ 95\% CL with P-ACT from $N_{\rm idr} < 0.379$ 95\% CL using \emph{Planck} alone. In this sense, the ACT DR6 data points us to a subset of the models which, so far, have been shown to be partially successful at addressing the Hubble tension \cite{Schoneberg:2021qvd,Khalife:2023qbu}.

The imminent release of the latest analysis of data from the SPT collaboration, which previously showed good agreement with {\it Planck}, will play a key role in determining whether EDE remains a viable model when attempting to address the Hubble tension. 

\label{sec:concl}

\vspace{2\baselineskip}
\textbf{Acknowledgments} --
We thank Boris Bolliet for help with the EDE emulator. We thank Adam Riess and Marc Kamionkowski for comments on an earlier version of this manuscript.
We thank Thibault Louis and Hidde Jense for help plotting the ACT DR6 residuals. TLS is supported by NSF Grants
No.~2009377 and No.~2308173. 
RC is funded by the Czech Ministry of Education, Youth and Sports (MEYS) and European Structural and Investment Funds (ESIF) under project number CZ.02.01.01/00/22\_008/0004632.
TS and VP acknowledges the European Union’s Horizon Europe research and innovation programme under the Marie Skłodowska-Curie Staff Exchange grant agreement No 101086085 – ASYMMETRY.
This work received funding support from the European Research Council (ERC) under the European Union’s HORIZON-ERC-2022 (grant agreement no. 101076865).
We gratefully acknowledge support from the CNRS/IN2P3 Computing Center (Lyon - France) for providing computing and data-processing resources needed for this work.

\appendix
\section{Summary tables}
\label{app:tables}
We report credible intervals for all parameters in Tab.~\ref{tab:CI} in the various analyses performed in this work.
We list individual $\chi^2$ contribution for all data sets in Tab.~\ref{tab:LCDM_chi2} for the $\Lambda$CDM model and Tab.~\ref{tab:EDE_chi2} for the EDE model.

\begin{table*}[]
\centering
\resizebox{\textwidth}{!}{
    \begin{tabular}{|c|c|c|c|c|c|c|}
    \hline

     &   \multicolumn{2}{c}{NPIPE-LS}&   \multicolumn{2}{|c|}{P-ACT-LS} &   \multicolumn{2}{|c|}{P-ACT-LBS}\\
         \hline

        SH0ES prior? & no & yes & no & yes & no & yes \\
\hline
 $100h$
	 & $67.96(68.45)^{+0.51}_{-0.93}$
	 & $71.65(71.96)\pm 0.81$
	 & $68.68(69.76)^{+0.62}_{-1.2}$
	 & $72.11(72.12)\pm 0.79$
	 & $69.71(70.98)^{+0.64}_{-1.3}$
	 & $72.34(72.49)\pm 0.72$
	 \\
$f_{\rm ede}(z_c)$
	 & $ < 0.065(0.043)$
	 & $0.113(0.122)\pm 0.022$
	 & $ < 0.092(0.075)$
	 & $0.127(0.134)^{+0.024}_{-0.020}$
	 & $ < 0.109(0.0902)$
	 & $0.126(0.133)\pm 0.021$
	 \\
$\log_{10}z_c$
	 & $3.62(3.82)^{+0.27}_{-0.30}$
	 & $3.616(3.574)^{+0.070}_{-0.14}$
	 & $3.54(3.55)^{+0.13}_{-0.22}$
	 & $3.548(3.554)^{+0.039}_{-0.033}$
	 & $3.54(3.549)^{+0.11}_{-0.18}$
	 & $3.550(3.563)\pm 0.043$
	 \\
$\theta_i$
	 & $--(2.92)$
	 & $2.69(2.817)^{+0.26}_{+0.026}$
	 & $--(2.65)$
	 & $2.59(2.690)^{+0.19}_{-0.039}$
	 & $--(2.708)$
	 & $2.67(2.716)^{+0.12}_{-0.082}$
	 \\
     \hline
$\Omega_\mathrm{c}h^2$
	 & $0.1221(0.1237)^{+0.0013}_{-0.0027}$
	 & $0.1299(0.1311)\pm 0.0027$
	 & $0.1231(0.1272)^{+0.0018}_{-0.0039}$
	 & $0.1323(0.1334)\pm 0.0031$
	 & $0.1226(0.1272)^{+0.0021}_{-0.0049}$
	 & $0.1315(0.1322)\pm 0.0029$
	 \\
$\Omega_\mathrm{b}h^2$
	 & $0.02228(0.02235)^{+0.00015}_{-0.00018}$
	 & $0.02264(0.02253)\pm 0.00020$
	 & $0.02259(0.02264)^{+0.00015}_{-0.00018}$
	 & $0.02286(0.02279)\pm 0.00016$
	 & $0.02268(0.02273)^{+0.00014}_{-0.00017}$
	 & $0.02285(0.02285)\pm 0.00016$
	 \\
$10^9A_\mathrm{s}$
	 & $2.114(2.118)^{+0.022}_{-0.025}$
	 & $2.157(2.156)^{+0.022}_{-0.026}$
	 & $2.126(2.145)\pm 0.024$
	 & $2.160(2.159)^{+0.021}_{-0.025}$
	 & $2.142(2.150)^{+0.023}_{-0.026}$
	 & $2.164(2.163)\pm 0.023$
	 \\
$n_\mathrm{s}$
	 & $0.9669(0.9715)^{+0.0045}_{-0.0067}$
	 & $0.9882(0.9878)\pm 0.0060$
	 & $0.9753(0.9795)^{+0.0055}_{-0.0070}$
	 & $0.9907(0.9894)\pm 0.0052$
	 & $0.9805(0.9853)^{+0.0055}_{-0.0069}$
	 & $0.9917(0.9912)\pm 0.0050$
	 \\
$\tau_\mathrm{reio}$
	 & $0.0574(0.0575)^{+0.0052}_{-0.0061}$
	 & $0.0609(0.0586)^{+0.0057}_{-0.0065}$
	 & $0.0585(0.0582)^{+0.0053}_{-0.0061}$
	 & $0.0597(0.0589)^{+0.0052}_{-0.0064}$
	 & $0.0625(0.0608)^{+0.0055}_{-0.0061}$
	 & $0.0614(0.0597)^{+0.0054}_{-0.0062}$
	 \\
$\Omega_\mathrm{m}$
	 & $0.3127(0.3127)\pm 0.0062$
	 & $0.2971(0.2979)\pm 0.0057$
	 & $0.3090(0.3094)\pm 0.0060$
	 & $0.2985(0.3015)\pm 0.0055$
	 & $0.2989(0.2989)\pm 0.0036$
	 & $0.2949(0.2963)\pm 0.0034$
	 \\

\hline
$\chi^2 (\rm EDE) - \chi^2(\Lambda CDM)$ & $-0.9$& $-30.7$&$-5.0$ & $-36.5$& $-7.0$& $-35.4$ \\
\hline
$Q_{\rm DMAP}$ & \multicolumn{2}{c|}{$3.5\sigma$} &  \multicolumn{2}{c}{$2.6\sigma$} &  \multicolumn{2}{|c|}{2.0$\sigma$} \\
        \hline
    \end{tabular}
 }
    \caption{Credible interval and best fit values (in parentheses) in the EDE model reconstructed from analyses of Planck PR4 (NPIPE), ACT DR6 combined with Planck (P-ACT),  CMB lensing (L), Pantheon-plus (S) and DESI DR2 BAO (B). We report results with and without  the SH0ES prior on $M_b$. For
parameters with two-sided constraints, we report the mean and 1$\sigma$ errors. For parameters with one-sided contraints, we report
the 2$\sigma$ limits. }
    \label{tab:CI}
\end{table*}

\begin{table*}[h!]
\centering
\begin{tabular}{l|rr|rr|rr}
\hline
\multicolumn{7}{c}{$\Lambda$CDM} \\
\hline
Data & \multicolumn{2}{c|}{NPIPE-LS} & \multicolumn{2}{c|}{P-ACT-LS}  & \multicolumn{2}{c}{P-ACT-LBS} \\
\hline
planck\_NPIPE\_highl\_CamSpec.TTTEEE & 10541.3 & 10545.6 & --- & --- & --- & --- \\
act\_dr6\_cmbonly.PlanckActCut       & ---     & ---     & 221.4 & 221.4 & 220.9 & 222.8 \\
act\_dr6\_cmbonly.ACTDR6CMBonly      & ---     & ---     & 157.5 & 160.6 & 159.5 & 161.1 \\
act\_dr6\_lenslike.ACTDR6LensLike    & 19.8    & 20.0    & 19.7  & 20.0  & 19.8  & 19.9  \\
planck\_2018\_lowl.EE\_sroll2        & 390.0   & 391.5   & 390.1 & 391.2 & 391.2 & 391.7 \\
planck\_2018\_lowl.TT                & 23.8    & 22.7    & 22.6  & 21.9  & 21.9  & 21.3  \\
sn.pantheonplus                      & 1403.6  & 1406.4  & 1403.9 & 1406.8 & 1405.7 & 1407.6 \\
bao.desi\_bao\_all                   & ---     & ---     & ---   & ---   & 12.6  & 10.4  \\
H0.riess2022Mb                       & ---     & 34.2    & ---   & 31.5  & ---   & 29.2  \\
\hline
$\chi^2_{\rm tot}$  & 12378.5 & 12420.4 & 2215.2 & 2253.4 & 2231.6 & 2264.0 \\

\hline
\end{tabular}
\caption{$\chi^2_{\rm min}$ contribution from each data set in the $\Lambda$CDM model.
\label{tab:LCDM_chi2}}
\end{table*}

\label{app:chi2}
 \begin{table*}[h!]
\centering
\begin{tabular}{l|rr|rr|rr}
\hline
\multicolumn{7}{c}{EDE} \\
\hline
Data & \multicolumn{2}{c|}{NPIPE-LS} & \multicolumn{2}{c|}{P-ACT-LS}  & \multicolumn{2}{c}{P-ACT-LBS} \\
\hline
planck\_NPIPE\_highl\_CamSpec.TTTEEE & 10540.6 & 10547.4 & --&-- & -- &--\\
act\_dr6\_cmbonly.PlanckActCut & --&-- & 217.7 & 218.3 & 217.7 &218.6\\

act\_dr6\_cmbonly.ACTDR6CMBonly & --&-- & 155.7 & 157.7 & 157.2 &158.7\\
act\_dr6\_lenslike.ACTDR6LensLike & 20.3 & 20.8 & 20.3 & 20.96 &  20.0&20.5 \\
planck\_2018\_lowl.EE\_sroll2 & 390.2 & 390.4 & 390.3 & 390.5 & 390.6& 390.5 \\
planck\_2018\_lowl.TT & 22.5 & 21.3 & 21.8 & 21.1 & 21.2&21.0  \\
sn.pantheonplus & 1404.0 & 1406.6 & 1404.4 & 1405.8 & 1406.3&  1406.9\\
ba.desi\_bao\_all & --&-- & --& --& 11.6 & 10.9\\
H0.riess2022Mb & --& 3.2 &-- & 2.5 &--& 1.5 \\
\hline 
$\chi^2_{\rm tot}$ & 12377.6 & 12389.7 & 2210.2 & 2216.9 & 2224.6 & 2228.6 \\

\hline
\end{tabular}
\caption{$\chi^2_{\rm min}$ contribution from each data set in the EDE model.}
\label{tab:EDE_chi2}
\end{table*} 

\section{DR4 vs DR6}
\label{app:DR4}

\begin{figure*}[!t]
    \centering
    \includegraphics[width=2\columnwidth]{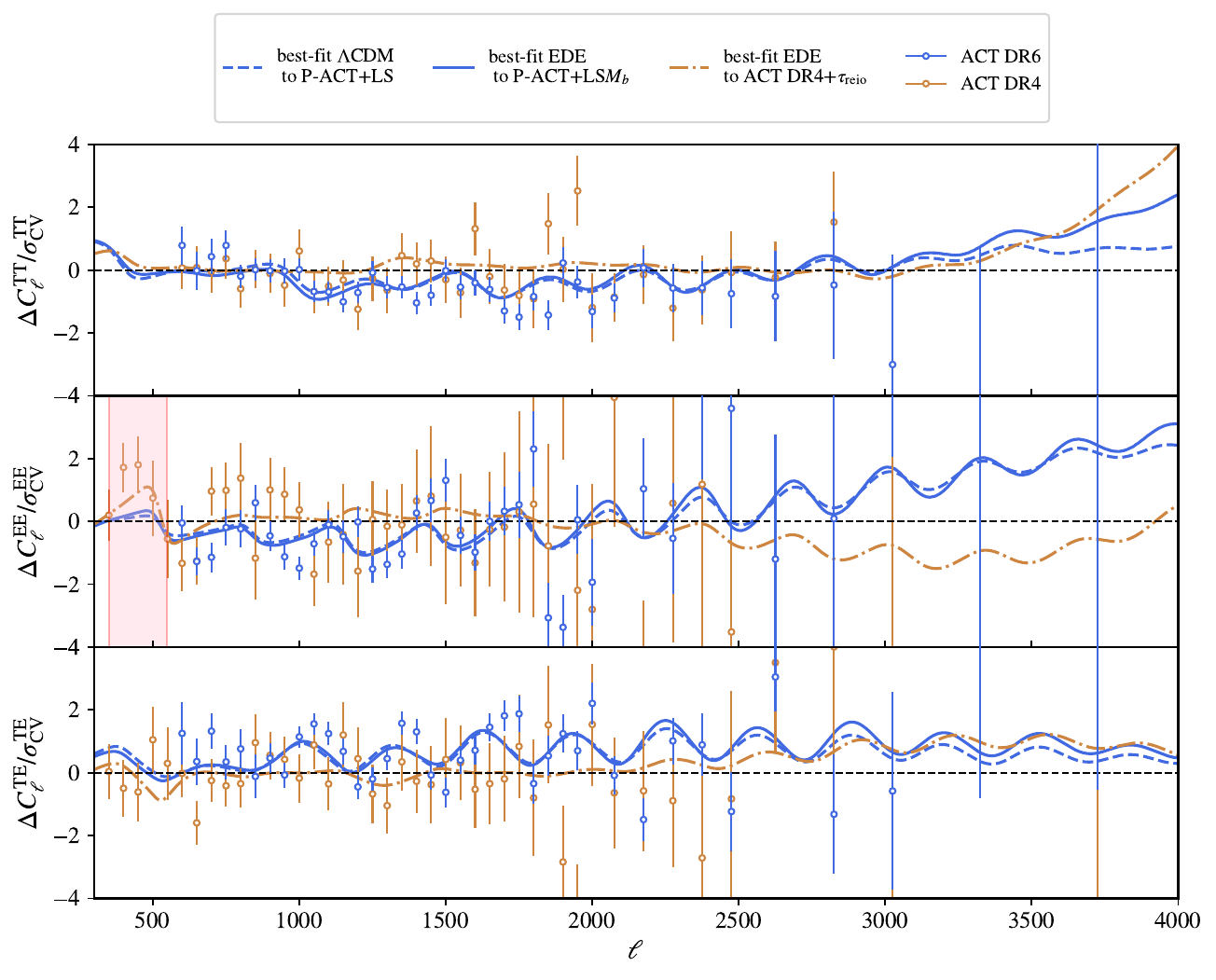}
    \caption{The best fit $\Lambda$CDM and EDE residuals with respect to the best fit $\Lambda$CDM model to ACT DR4 + $\tau_{\rm reio}$  \cite{Hill:2021yec} normalized by the cosmic variance (CV) limits for each power spectrum (note that the data may have uncertainties smaller than the CV limit since they are binned). We show residuals for  the best fit $\Lambda$CDM model using P-ACT+LS (dashed), the best fit $\Lambda$CDM model using ACT DR4 + a prior on $\tau_{\rm reio}$ (dotted), while for EDE we use P-ACT+LS$M_b$ (solid) and ACT DR4 + a prior on $\tau_{\rm reio}$ (dot dashed). The ACT DR4 residuals were computed using parameters given in Ref.~\cite{Hill:2021yec}. Both the ACT DR4 and DR6 bandpowers are marginalized over foreground and other nuisance parameters (i.e. the `lite' versions of their respective likelihoods) \cite{ACT:2020gnv,ACT:2025fju}. These bandpowers are not statistically independent, especially at $\ell \gtrsim 2000$ where the foreground model leads to strong correlations. }
    \label{fig:residuals}
\end{figure*}
The shift in posterior distributions for EDE going from ACT DR4 to DR6 can be better understood by comparing their respective bandpowers. Fig.~\ref{fig:residuals} shows these bandpowers as well as the best fit residuals of $\Lambda$CDM fit to P-ACT+LS and EDE fit to P-ACT+LS$M_b$ and to ACT DR4 + a prior on $\tau_{\rm reio}$ \cite{Hill:2021yec} with respect to the $\Lambda$CDM best fit to ACT DR4 + a prior on $\tau_{\rm reio}$ \cite{ACT:2020gnv,Hill:2021yec}. 

First we note that the lowest multipole bins in the polarization data from ACT DR4 -- in particular for the EE power spectrum-- were identified as driving the preference for non-zero $f_{\rm EDE}(z_c)$ when using ACT DR4 + a prior on $\tau_{\rm reio}$ \cite{Hill:2021yec}.  The removal of these bins from the DR6 data release leaves open the question of how EDE may have fit the new data had they been present. However, it is important to note that, when the ACT DR4 data are combined with other CMB and LSS data sets, the preference for non-zero $f_{\rm EDE}(z_c)$ persists while the role that the DR4 low multipole EE power spectrum plays is significantly reduced \cite{Poulin:2021bjr,Smith:2022hwi}. 

Fig.~\ref{fig:residuals} shows that when going from ACT DR4 to DR6 we not only see a significant decrease in the size of the error bars but also a systematic shift in the means so that the TT and EE power spectra are lower and TE power spectrum is higher in DR6. In addition, the residuals indicate that there is a slight phase shift between the best fit models to the two data releases.

\section{Profile likelihoods}
\label{app:profile}
\begin{figure}
    \centering
    \includegraphics[width=1\columnwidth]{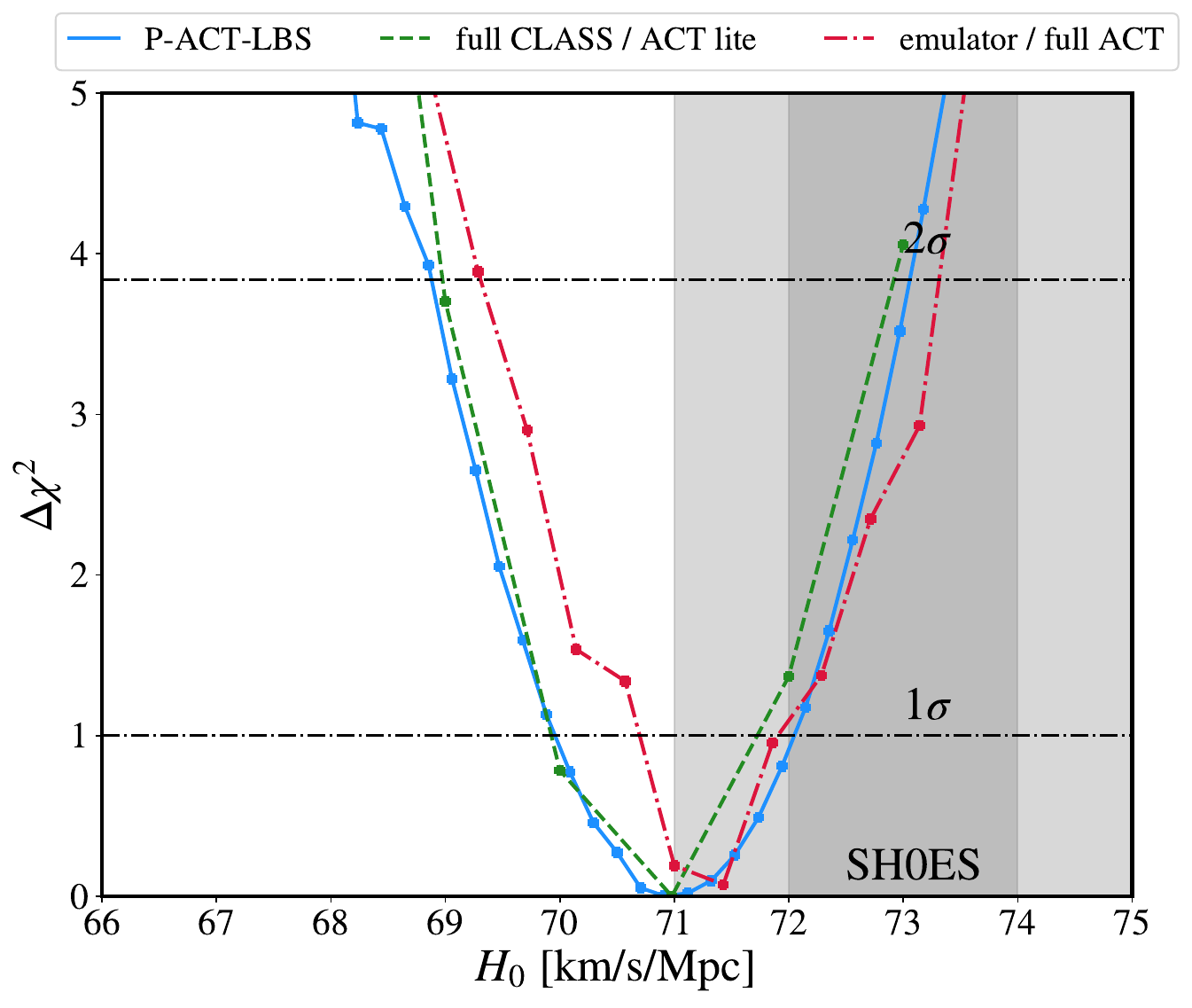}
\caption{Profile likelihood of $H_0$ for our baseline analysis that uses the EDE emulator developed in Ref.~\cite{Qu:2024lpx} and ACT  \texttt{lite} data, compared with the profile likelihood reconstructed from the use of the code \texttt{class\_ede} \cite{Hill:2020osr}  when analyzing the same datasets, and the use of the emulator when analyzing the full ACT \texttt{mf-like} likelihood.  }
    \label{fig:H0_prof_test}
\end{figure}

We show in Fig.~\ref{fig:H0_prof_test} the profile likelihood of $H_0$ for our baseline analysis that uses the EDE emulator developed in Ref.~\cite{Qu:2024lpx} and ACT \texttt{lite} data, compared with the profile likelihoods reconstructed from the use of the code \texttt{class\_ede} \cite{Hill:2020osr} when analyzing the same datasets, as well as the use of the emulator when analyzing the full ACT \texttt{mf-like} likelihood. All analyses include the `P-ACT' data, DESI DR2, lensing and Pantheon-plus (uncalibrated). 
As the full code is much slower due to the necessary increase in theoretical precision, we run a very sparse profile at five values of $H_0\in[68,73]$ together with a search for the global best-fit. When using full ACT data with the emulator, we run instead at fifteen values of $H_0$ within the same range.
Though there are some fluctuations in the profiles, all analyses lead to similar best-fit models and confidence level. In particular, they all allow for $H_0 = 73$ km/s/Mpc at $\sim 2\sigma $. Note that the full data appears to disfavor $\Lambda$CDM slightly more than the \texttt{lite} data. 

\bibliographystyle{apsrev4-1}
\bibliography{bib} 

\end{document}